# Inductance Meets Memory in the Quantum Magnet $Mn_3Si_2Te_6$

Tristan R. Cao[1], Gabriel Schebel[1], Arabella Quane[1], Hengdi Zhao[1], Yu Zhang[1], Feng Ye[2], Longji Cui[3], and Gang Cao[1*]

[1]Department of Physics, University of Colorado at Boulder, Boulder, CO 80309, USA

[2]Neutron Scattering Division, Oak Ridge National Laboratory, Oak Ridge, Tennessee 37831, USA

[3]Department of Mechanical Engineering, University of Colorado at Boulder, Boulder, CO 80309, USA

*Corresponding author: gang.cao@colorado.edu

Orbital degrees of freedom offer a largely untapped route to emergent dynamical phenomena in correlated quantum materials. However, it remains unclear whether collective orbital states can intrinsically generate both reactive and memory functionalities in a bulk system. Here we show that in the ferrimagnet $Mn_3Si_2Te_6$, nonequilibrium reconfiguration of chiral orbital currents produces both emergent inductance and nonvolatile memristance as intrinsic properties of a single crystal. At low frequency and under a magnetic field along the c axis, coherent orbital-current domains generate robust clockwise inductive I-V loops. At higher frequency and low field, current-driven first-order reconfiguration leads to incomplete reversal and metastable trapping, producing an intrinsic electromotive force and a finite remanent voltage at zero current. These results establish orbital currents as a class of quantum state variables that encode both reactive and memory functionalities, opening routes toward intrinsically reconfigurable and energy-efficient electronic systems.

**Introduction**

Chiral orbital currents (COC) have recently emerged as an organizing principle in correlated quantum materials [1-10]. In our earlier studies [1-4], we have demonstrated that COC in the ferrimagnet $Mn_3Si_2Te_6$ drive a wealth of unconventional transport phenomena, including colossal magnetoresistance (CMR) [1, 2], first-order bistable switching [1], an unusual current-sensitive Hall effect, etc. [3]. These findings established orbital currents as a bona fide degree of freedom beyond spin and charge, opening the door to an "orbitronic" framework for understanding and enginnering transport. It is worth noting that the slow dynamics of the COC reconfigurations, spanning seconds to minutes, is a defining characteristic of the orbitronics because of the rigid coupling between the lattice and orbital currents extending over multiple atomic sites [1–3]. This further distinguishes COC from conventional fast electronic processes, favoring emerging applications where stability, tunability, and retention are prioritized [1]. More recently, we have demonstrated that coherent COC domains can serve as an intrinsic source of inductance [5]: When stabilized by a magnetic field, H, along the hard *c* axis (H || *c*), $Mn_3Si_2Te_6$ exhibits robust clockwise inductive I-V loops at low frequencies *f* < 6 Hz, meaning that decreasing current, *I*, leads to an increase in voltage, *V* (Figs.1a-1b), a striking reversal of the counterclockwise I-V behavior expected for conventional resistive, magnetic or capacitive systems [5]. This discovery identifies orbital-current dynamics as a mechanism for emergent inductance in a bulk crystal, distinct from the geometric, kinetic and other inductance known in existing materials and devices [11-16]. In correlated quantum materials, hysteretic and history-dependent transport responses often provide the most direct experimental window into slow collective orbital and magnetic degrees of freedom that remain hidden in linear probes. Such intrinsic history-dependent responses are of interest not only for reconfigurable electronics but also as probes of nonequilibrium collective quantum states.

In this work, we uncover a qualitatively different regime of behavior, arising from a competition between partially coherent and disordered orbital textures, that hosts both inductors and nonvolatile memristors in a bulk material (Figs.1b-1c). At zero or low H || c and frequencies $f$ > 6 Hz (a crossover range $f_c$ ~ 6 Hz marks the onset of nonequilibrium response), $Mn_3Si_2Te_6$ exhibits both inductive transport and a nonvolatile memristive response, marked by a sizable remanent voltage at zero current, $V_m \neq 0$ at $I = 0$ (contrasting to volatile memristors in which $V = 0$ at $I = 0$ [17]). Unlike nanoscale oxide memristors, where memory typically originates from ionic migration, filament formation, or interfacial redox processes, the present behavior emerges intrinsically in a homogeneous bulk single crystal, eliminating common device-related artifacts and highlighting a collective orbital origin. As shown in Fig.1c, the I-V loop in this regime is clearly clockwise, demonstrating intrinsic inductive character [5], and retains memory of prior bias states, establishing a bulk nonvolatile memristor. These effects vanish and the I-V loop becomes counterclockwise when $\mu_0 H_{||c} \geq 3$ T, where coherent COC begin to form [1, 3, 5] and the competition between coherent and disordered COC vanishes. However, the memristive behavior remains largely insensitive to H || a at which COC do not form [1], underscoring a distinct orbital origin.

In short, Regime A corresponds to conditions where coherent COC domains are stabilized at high H || c and $f < f_c$ ~ 6 Hz, whereas Regime B corresponds to a nonequilibrium regime of disordered or partially coherent orbital textures at zero or low H || c and $f > f_c$ ~ 6 Hz where inductance and nonvolatile memory coexist.

**Results and Discussion**

Before presenting detailed data, we first turn to the microscopic origin of emergent inductance and nonvolatile memory from orbital texture dynamics. As illustrated schematically in **Fig. 1c**, the clockwise I-V loops and nonvolatile remanent voltage $V_m$ arise because current-driven, first-order reconfiguration of COC textures (① → ② in Fig.1c) produces a time-dependent orbital magnetization $M_O(t)$ that cannot fully reverse on decreasing current, thereby generating an intrinsic electromotive force (*emf* ∝ *-dI/dt*) and metastable memory (③ in Fig.1c).

Specifically, the clockwise I-V loops and nonvolatile remanent voltage arise from the nonequilibrium dynamics of COC textures, rather than from conventional electronic, magnetic, or ionic mechanisms. In this material, COC form extended, multi-site loop patterns that are rigidly coupled to the lattice and magnetic background, resulting in unusually slow, collective dynamics [1,5].

When an external current *I* is applied, the electronic system is driven out of equilibrium and the orbital textures reorganize in response [5]. Importantly, this reconfiguration does not occur continuously. Instead, the system evolves through a sequence of first-order transitions between nearly degenerate orbital configurations [1, 5], separated by discrete energy barriers. These barriers are evidenced experimentally by reproducible threshold voltages ($V_{R1}$, $V_{R2}$, $V_{R3}$) (Fig.1c and discussion below), which mark abrupt changes in the differential resistance (R) and signal collective, domain-wide rearrangements of orbital currents.

On increasing current *I*, the orbital textures discretely expand and partially align into more coherent COC configurations, reducing R, which is manifest at $V_{R1}$, $V_{R2}$, $V_{R3}$ (Fig.1c). However, when the current *I* is subsequently decreased, these reorganized COC cannot fully retrace their path across the same energy barriers within a single cycle because the reorganized COC are realized via a series of first-order transitions evidenced by $V_{R1}$, $V_{R2}$, $V_{R3}$ (Fig.1c). This incomplete reversal of orbital-current circulation produces a time-dependent orbital magnetization $M_O(t)$ and an associated *emf*, analogous to but microscopically distinct from Faraday induction. The resulting *emf* ($\propto - dI/dt$) opposes the change in current giving rise to a clockwise inductive I-V loop**,** where the voltage V increases as the current I decreases (Fig.1c).

At low frequencies (*f* < 6 Hz) and in the presence of H || *c*, the system has sufficient time or energetic bias to relax toward a unique coherent orbital configuration. In this regime (Regime A: low *f* < 6 Hz + high H || *c* in Fig.1a), the induced *emf* is transient and vanishes once the current cycle completes, yielding inductive behavior without memory.

In contrast, at higher frequencies and zero or low H || *c*-axis (Regime B: higher *f* > 6 Hz + zero or low H || *c*. in Fig.1c), the current oscillates faster than the relaxation time of the orbital textures with characteristic timescales on the order of tens of milliseconds, as suggested by the crossover near $f_c \sim$ 6 Hz (Fig.1b). Under these conditions, partial reconfiguration during the rising current branch becomes metastably trapped when the current decreases. This trapping leaves a persistent orbital imbalance that manifests as a finite remanent voltage at zero current, i.e., $V_m \neq 0$ at $I = 0$ (Fig.1c). The system thus retains a memory of its prior dynamical state, establishing a nonvolatile memristive response with a lifetime exceeding tens of milliseconds.

The coexistence of inductance and nonvolatile memory therefore reflects the same underlying physics: Slow, first-order reconfiguration of collective orbital textures across a multi-minima free-energy landscape. The frequency at which $V_m$ is maximized corresponds to an optimal dynamical window in which the drive is fast enough to induce barrier crossing but slow enough to allow substantial orbital reorganization within each cycle (Fig.1b). This behavior sharply distinguishes the existing mechanism from conventional electronic or ionic memristors [18-20] and identifies COC as an emergent state variable capable of encoding both reactive and memory functionalities.

In the following, we demonstrate the detailed inductive and memristive behaviors of bulk single crystals of $Mn_3Si_2Te_6$ as functions of frequency, I-V cycle, current, magnetic field, and temperature. It needs to be pointed out that our thorough thermal diagnostics conclusively rule out Joule heating, a gradual, diffusive, and isotropic process, as the origin of these rare properties. Supporting data and discussion are provided in Supplementary Figures, SFigs. 1-13 and Supplementary Note 3 and have been similarly addressed in our prior studies [1, 3, 5]. Transport measurements were done using the four-probe technique (Fig.1b).

$Mn_3Si_2Te_6$, first reported in 1986 [21], crystallizes in the trigonal space group *P-31c* (No. 163), forming a trimer-honeycomb lattice (Fig.1b) [4]. It is an electrical insulator and orders ferrimagnetically below $T_C$ = 78 K, adopting a noncollinear magnetic structure (Fig.1b) [4, Supplementary Note 1]. Owing to its unusual orbital characteristics, $Mn_3Si_2Te_6$ has been the subject of extensive studies in recent years [1-6, 22-27].

Frequency Dependence and Crossover to Memristive Behavior at H = 0. We now examine the frequency dependence of the zero-field response. At low $f < f_c$ ~ 6 Hz and H = 0, the I-V curves are counterclockwise hysteresis loop (Fig.2a). This behavior is common in magnetic materials and electronic systems [19, 27-29] and is discussed in our earlier work [5]. For $f \geq f_c$, the I-V loops open into pronounced clockwise shapes, indicating the occurrence of inductance (Fig. 2c). The inductive behavior is particularly strong for $f$ = 24.4 Hz at which the voltage V increases as the current I decreases (Fig. 2b).

Most strikingly, a finite remanent voltage $V_m$ emerges at $I = 0$, confirming the onset of a nonvolatile memristor state. For example, for $f$ = 24.4 Hz and $T$ = 10 K, $V_m$ = 2.9 V at $I = 0$ (Fig.2b). This behavior can be described by $V = IR + LdI/dt$, where R and L are resistance and inductance, respectively. Since $V_m = LdI/dt$ at $I = 0$, $L \approx V_m(2\pi f I_{max})^{-1}$ = 0.76 Henry for $f$ = 24.4 Hz, $V_m$ = 2.9

V and $I_{max}$ = 25 mA. This is an enormous value for a millimeter-sized crystal, far beyond geometric inductance (typically nanohenries) [11-16]. The extracted $V_m$ from Fig. 1c reaches a pronounced maximum near $f$ = 24 Hz (Fig. 2d). At still higher frequencies, the current oscillates too rapidly for the orbital domains to fully reconfigure within each half-cycle, causing $V_m$ to roll off. Clearly, the peak in $V_m$ near 24 Hz reflects the frequency at which orbital reconfiguration and metastable trapping are maximized whereas the crossover at $f_c \sim$ 6 Hz, evident in Figs. 2a and 2c, marks the onset of nonequilibrium response. These features reflect slow collective orbital reconfiguration on timescales of order tens of milliseconds.

Multi-Sweep I-V Curves Revealing the System Retaining Memory of the Prior Cycle. The multi-sweep current-driven I-V curves provide insight into the memory lifetime of the orbital texture. At $f$ = 24.4 Hz (Fig. 3a), successive sweeps exhibit clear state carry-over: Even when the next sweep begins at $I_o$ = 0, a finite initial remanent voltage $V_o \neq 0$ is present, directly reflecting a persistent emf from the prior cycle. This demonstrates that partially reconfigured COC domains remain trapped between sweeps, acting as an internal *emf* source on the next sweep, with a relaxation time longer than the drive period. In other words, the partial relaxation of the orbital texture toward equilibrium, but not instantaneously, indicates a finite memory lifetime at $f$ = 24.4 Hz (see SFig.1 for more data). In contrast, at $f$ = 0.3 Hz (Fig. 3b), no such carry-over is observed; the system fully relaxes within the ~ 1.7 s half-cycle, eliminating the initial remanent voltage $V_o$. All in all, these data in Fig. 3 bracket the orbital memory lifetime to lie between tens and hundreds of milliseconds. Note that the data in Fig. 3 are collected from a different crystal to confirm reproducibility of the observed phenomena.

Current-Driven Reconfiguration and Discrete Voltage Thresholds. Indeed, the I-V curves undergo abrupt slope changes at well-defined threshold voltages $V_{R1}$ and $V_{R2}$ as $I$ rises (Figs. 1c and 2b), and for current limits $I_{max} > 40$ mA, an additional threshold voltage $V_{R3}$ emerges, as shown in Fig. 4a. These threshold-like features mimic the knee voltage required for a diode to begin conducting significant current. The three reproducible threshold/knee voltages (e.g., $V_{R1} = 0.9$, $V_{R2} = 2.6$, $V_{R3} = 3.0$ V at $f = 24.4$ Hz, H = 0 and $T = 10$ K) indicate that the COC reorganize through discrete threshold events rather than continuously. These steps likely correspond to crossing discrete energy barriers separating metastable orbital configurations, consistent with a multi-minima free-energy landscape and supporting the picture of first-order COC reconfigurations that underpin the observed inductive and memristive responses. It is thus natural that the I-V curves follow the Ohm's law between $V_{R1}$, $V_{R2}$ and $V_{R3}$ where no COC reconfigurations occur (Fig.4a).

Converting the I-V curve at $I_{max} = 90$ mA (Fig. 4a) into differential resistance $dV/dI$ versus current I yields Fig. 4b, which directly visualizes the evolution of $dV/dI$ with discrete COC reconfigurations. As I increases, $dV/dI$ decreases nonlinearly, with pronounced drops near the onsets of *R1*, *R2*, and *R3*, confirming first-order, successive expansions of COC domains toward more coherent COC configurations. Upon decreasing *I*, these reorganized COC domains cannot retrace their path or reverse the first-order COC reconfigurations, leading to the generation of a sizable *emf* (Fig. 4a) and a rapid increase in $dV/dI$ (Fig. 4b). This hysteretic behavior reinforces the view that the emergent inductance and memory stem from current-driven, first-order reconfigurations of orbital textures across discrete energy barriers.

Magnetic-Field Dependence and Orientation Selectivity. We now turn to the effects of magnetic field orientation on the inductance and nonvolatile memristor regime at the representative $f = 24.4$ Hz and 10 K. For H || *c*, both the clockwise inductive I-V loops and $V_m$ are progressively suppressed, vanishing once $\mu_0 H_{||c} \geq 1.5$ T; further increase H || *c*, e.g., $\mu_0 H_{||c} = 3$ T, reverses the I-V loops from clockwise to counterclockwise, as shown in Fig. 5a. This is precisely the field scale

at which coherent COC domains begin to form, resulting in CMR [1, 5]. It indicates that the competition between coherent and disordered COC responsible for memory is destabilized by fully field-aligned COC. Conversely, for H ∥ $a$ at which no COC form, the memristive behavior and $V_m$ remain well-defined up to $\mu_0H_{\parallel a}$ = 14 T, underscoring the persistent competition between coherent and disordered COC, which enables memory (Fig. 5b). This becomes more evident in SFig. 2 in which the field-dependence of $V_m(H)$ and $a$-axis magnetoresistance ratio $[\rho_a(H)-\rho_a(0)]/\rho_a(0)$ $(=\Delta\rho_a/\rho_a(0))$ is plotted for comparison ($\rho_a$ is the a-axis resistivity). That the field dependence of $V_m(H\|a)$ and $V_m(H\|c)$ closely follows that of $\Delta\rho_a/\rho_a(0)(H\|a)$ and $\Delta\rho_a/\rho_a(0)(H\|c)$, respectively, reinforces the interpretation that the observed inductance and nonvolatile memristance originate from orbital-current dynamics, and further highlights the competition between disordered and coherent COC states.

Finally, we examine the temperature dependence of the inductive and nonvolatile memristive response. While for $T < T_C$, robust clockwise loops with finite $V_m$ confirm the coexistence of inductive and nonvolatile behavior (Fig.5c and SFig. 3), it is striking that $V_m$ persists well above the $T_C$ = 78 K (SFig. 3b) and exhibits a modest peak near $T_C$ before gradually decreasing at higher temperatures, vanishing near 130 K (Fig. 5d). This persistence demonstrates that long-range COC and magnetic orders are not required for memory; instead, the effect originates from fluctuating or disordered orbital textures that remain nearly degenerate and slow to relax even in the paramagnetic state. The small enhancement of $V_m$ near $T_C$ can be attributed to critical slowing down of orbital and spin fluctuations, which increases the likelihood of metastable trapping. Above $T_C$, short-range COC fluctuations continue to sustain nonvolatile behavior, although with reduced amplitude. Indeed, our neutron study of $Mn_3Si_2Te_6$ reveals magnetic fluctuations persisting up to 200 K [4].

**Conclusions**

The data in Figs. 1-5 establish a comprehensive picture in which $Mn_3Si_2Te_6$ hosts two orbital regimes, A and B. In the field-stabilized Regime A, coherent COC domains yield emergent inductance without memory [5], whereas in the disordered Regime B, nearly degenerate orbital states driven out of equilibrium at higher frequency produce both inductance and nonvolatile memristance (Fig. 1c). The persistence of this behavior above $T_C$ (SFig.3, Fig.5d), as well as its field dependence (Figs.5a-5b), underscores that the underlying mechanism is rooted in orbital dynamics rather than static long-range order. These results demonstrate that orbital textures represent a bona fide electronic degree of freedom capable of realizing three of the four fundamental passive circuit elements [18], inductors and memristors as well as resistors, within a single bulk quantum material (no capacitive behavior is discerned (Supplementary Note 2 and SFig.4) [5].

Importantly, the emergent inductance is not a smooth relaxational response [5] but arises from first-order, successive reconfigurations of COC textures driven by the AC current (Figs.1c and 4). On the rising branch of the cycle, orbital domains reorganize and expand into more coherent configurations once discrete energy barriers are overcome. Upon decreasing current, these reorganized textures cannot fully retrace their path across the same barriers, leading to incomplete reversal of orbital circulation and the generation of an intrinsic *emf* that manifests as a clockwise inductive loop in the I-V curves. At $f > f_c \sim 6$ Hz the system is unable to fully relax within each cycle, leaving the orbital texture metastably trapped in nearly degenerate configurations and producing a finite remanent voltage $V_m$ at $I = 0$ zero, thereby establishing a nonvolatile memristive response with a memory lifetime on the order of 10-100 ms. The pronounced maximum in $V_m$ near $f \sim 24$ Hz identifies an optimal dynamical window where the drive is fast enough to induce barrier crossing yet slow enough to allow substantial orbital reconfiguration. At $f > 24$ Hz, the COC cannot reorganize efficiently, resulting in a reduced $V_m$; whereas at $f_c \sim 6$ Hz, the COC has enough time to relaxes toward equilibrium (Figs. 2c-2d).

This work reveals that complex collective order parameters in bulk quantum materials can act as intrinsic state variables that encode both reactive and memory functionalities. The coexistence of emergent inductance and nonvolatile memristance in a single crystal establishes a rare example in which circuit behavior emerges directly from correlated quantum state, rather than from engineered device components. These findings open a route toward intrinsically reconfigurable, ultra-low-power, and quantum-inspired electronic platforms where functionality is dynamically encoded in the bulk quantum matter.

**Methods**

Single crystals of $Mn_3Si_2Te_6$ were grown using a flux method. Measurements of crystal structures were performed using a Bruker Quest ECO single-crystal diffractometer with an Oxford Cryosystem providing sample temperature environments ranging from 80 K to 400 K. Chemical analyses of the samples were performed using a combination of a Hitachi MT3030 Plus Scanning Electron Microscope and an Oxford Energy Dispersive X-Ray Spectroscopy (EDX). The measurements of the electric resistivity and I-V characteristic were carried out using a Quantum Design (QD) Dynacool PPMS system having a 14-Tesla magnet and a set of external Keithley meters that provides current source and measures voltage with a high precision.

It is important to point out that the I-V curves presented in the main text are current-driven. I-V curves are taken using the QD Dynacool PPMS, whose electronics applies a triangular excitation waveform that always starts and ends at zero bias (see Manual for QD PPMS Electrical Transport Option).

In our experiments, the contact resistance was measured to be on the order of 10 Ω, at least 4 orders of magnitude smaller than the sample resistance in the insulating state (>$10^5$ Ω). Importantly, these measurements were made using standard four-probe configurations, with current and voltage leads separated to eliminate contact contributions to voltage measurements.

**Data availability**

The data that support the findings of this study are available on request from the corresponding author [GC].

**Acknowledgement**

This work is supported by U.S. National Science Foundation via Grant No. DMR 2204811.

**Author contributions**

T.R.C. conceived the idea of this work and conducted measurements of the physical properties and data analysis; G.S. synthesized the single crystals; A.Q. characterized the crystal structure of the samples; H.Z., Y.Z., F.Y. and L.C. contributed to valuable discussion and revisions to the manuscript; G.C. directed this work, conducted experiments of the physical properties and data analysis, and wrote the paper.

**Competing interests**

The authors declare no competing interests.

**Figure captions**

**Fig. 1. Two orbital regimes and field-frequency phase diagram. a, Regime A: Coherent chiral orbital currents (COC) at low frequency $f$ + high magnetic field H || $c$:** Representative I-V curve showing a characteristic clockwise inductive loop and schematic illustrating a two-stage process of ① formation of coherent COC acting as mesoscopic "coils" that generate ② inductance on current decrease [5]. **b, Field-frequency phase diagram** showing the crossover near crossover frequency $f_c$ ~ 6 Hz that separates Regimes A and B. The optimal energy-absorption window is near 24 Hz. **Inset:** A schematic showing the current and voltage leads on the sample and field orientations. **c, Regime B: Partially coherent and disordered COC at high f + zero/low H || $c$.** Representative I-V curve showing both a clockwise inductive loop and memristive response with remanent voltage $V_m \neq 0$ at current $I = 0$. Note that threshold voltages $V_{R1}$ and $V_{R2}$ signal discrete threshold events. The schematic illustrating a dynamic orbital process of ① disordered COC, ② partially disordered COC + partially coherent COC (blue) on current increase and ③ emergent inductance + trapping/memory in metastable COC states (cyan) on current reduction. Specifically, current-driven COC reconfiguration produces a time-dependent orbital magnetization $M_O(t)$ (②). Upon decreasing current $I$, incomplete reversal across discrete energy barriers generates an

electromotive force $emf \propto -dI/dt$, resulting in clockwise inductive I-V loops. At $f > f_c$, metastable trapping of the orbital configuration leaves a $V_m \neq 0$ at $I = 0$ ③.

**Fig.2. Frequency $f$ dependence and crossover to memristive behavior at magnetic field H = 0. a-c,** I-V loops measured at temperature $T = 10$ K and H = 0 for $f = 0.3$ Hz - 97.6 Hz. Counterclockwise I-V loops at crossover frequency $f_c < 6$ Hz (a), and clockwise I-V loops emerging with increasing $f$, and a clear remanent voltage $V_m$ marked by solid circles (b, c). **d,** $V_m$ vs $f$, peaking near 24 Hz, identifying an optimal energy-absorption window where current-driven first-order reconfiguration of domains of chiral orbital currents (COC) is most efficient. This resonance-like peak marks the $f$ range where partial COC reorganization and metastable trapping are maximized.

**Fig. 3. Multi-sweep I-V curves revealing the system retaining memory of the prior cycle. a,** Multi-sweep I-V curves at 24 Hz showing state carry-over. Successive voltage-driven sweeps (up to 4 sweeps) exhibit a finite initial remanent voltage $V_o \neq 0$ at current $I = 0$, marked by the solid squares, demonstrating that the system retains memory of the prior cycle. The partial relaxation of the orbital texture toward equilibrium, but not instantaneously, indicates a finite memory lifetime on the order of 10 – 100 ms. Note the clockwise inductive loops. **b,** Multi-sweep I-V curves at 0.3 Hz showing full relaxation. Successive two sweeps retrace without offset, and $V_o = 0$ at $I_o = 0$. The absence of carry-over demonstrates that the metastable orbital configuration relaxes fully within the ~1.7 s half-cycle at 0.3 Hz. Note the counterclockwise loops, typical of a normal state. The data in Fig.3 are collected from a different crystal to confirm reproducibility of the observed phenomena.

**Fig. 4. Current-driven reconfiguration and discrete voltage thresholds. a,** I-V loop at frequency $f = 24.4$ Hz showing distinct voltage steps of threshold voltages $V_{R1}$, $V_{R2}$ and $V_{R3}$

associated with overcoming discrete energy barriers for successive reconfigurations of chiral orbital currents (COC) including coherent COC states (blue) and metastable COC states (cyan). At current $I = 0$, remanent voltage $V_m \neq 0$ due to metastable trapping (cyan). **b**, Differential resistance *dV/dI* extracted from the I-V curve with the current limit $I_{max} = 90$ mA, showing sharp drops at the onsets of COC reconfigurations *R1, R2* and *R3* on current increase and rapidly rising *dV/dI* on current decrease. The hysteresis between increasing and decreasing current branches confirms incomplete reversal and memory formation.

**Fig. 5. Magnetic-field dependence and orientation selectivity of Regime B. a,** Evolution of I-V loops with increasing magnetic field H ∥ c at frequency $f = 24.4$ Hz: Both inductive loop area and remanent voltage $V_m$ decrease and vanish above ∼3 T as coherent COC form. Note that I-V loops are clockwise at $\mu_o H_{\|c} \leq 1.3$ T and becomes counterclockwise $\mu_o H_{\|c} = 3$ T. **b,** Field dependence with H ∥ a: I-V loops and remanent voltage $V_m$ remain largely insensitive to increasing H, indicating that only H∥c couples strongly to coherent COC. **c,** Temperature dependence: I-V loops at representative temperatures from 5 K to 130 K, H = 0, and $f = 24.4$ Hz, showing robust clockwise inductive loops and finite $V_m$ well into the paramagnetic state. **d**, $V_m(T)$ compared with the *a*-axis magnetization $M_a(T)$ at 0.1 T: A small peak near $T_C$ reflects critical slowing down of orbital and spin fluctuations, which enhances metastable trapping. The persistence of $V_m$ above $T_C$ implies the importance of orbital dynamics rather than static long-range order for nonvolatile memristance.

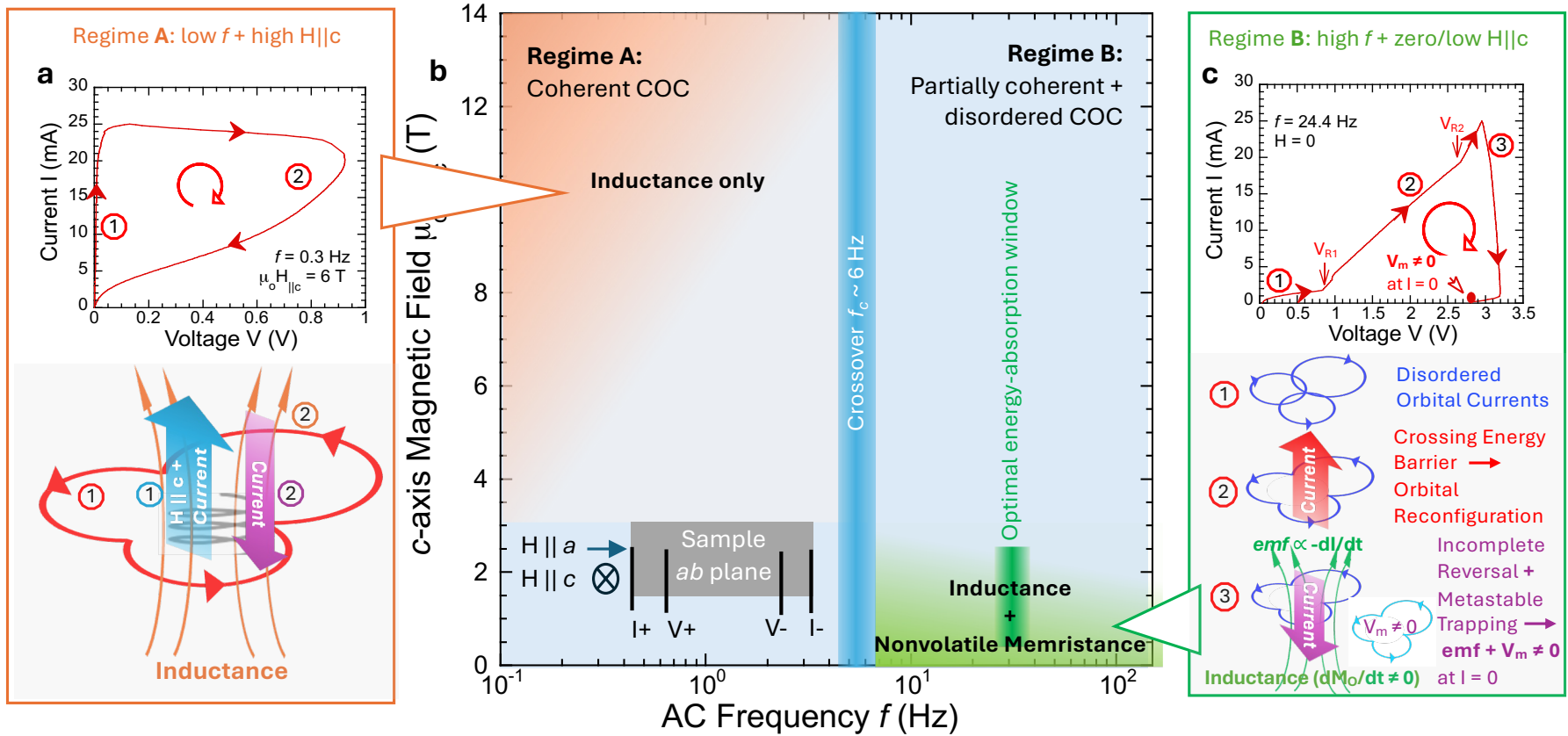

Figure 1

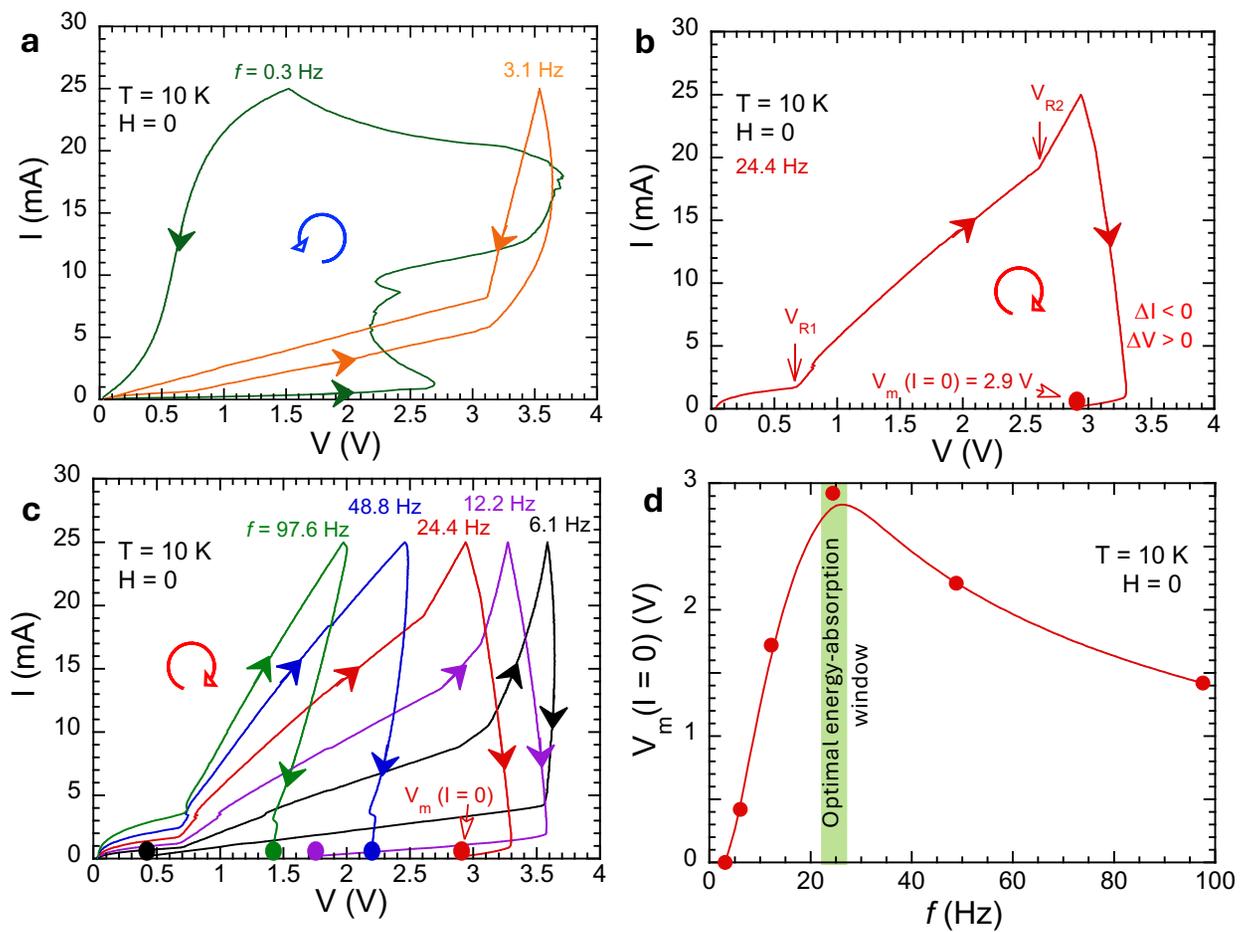

Figure 2

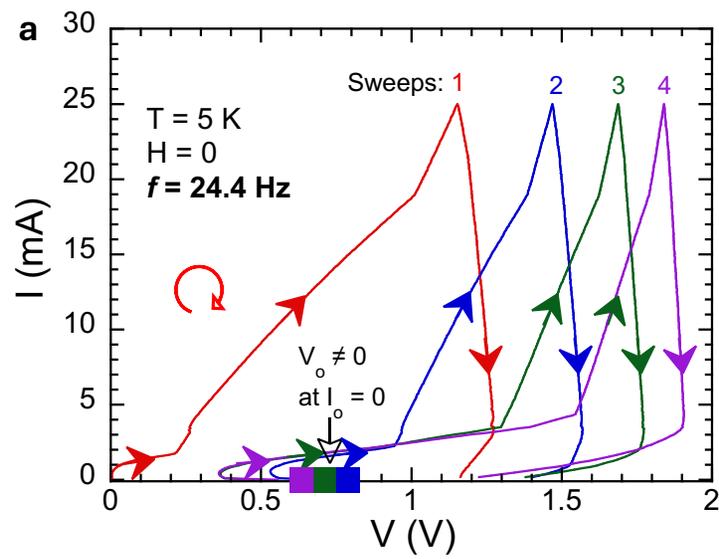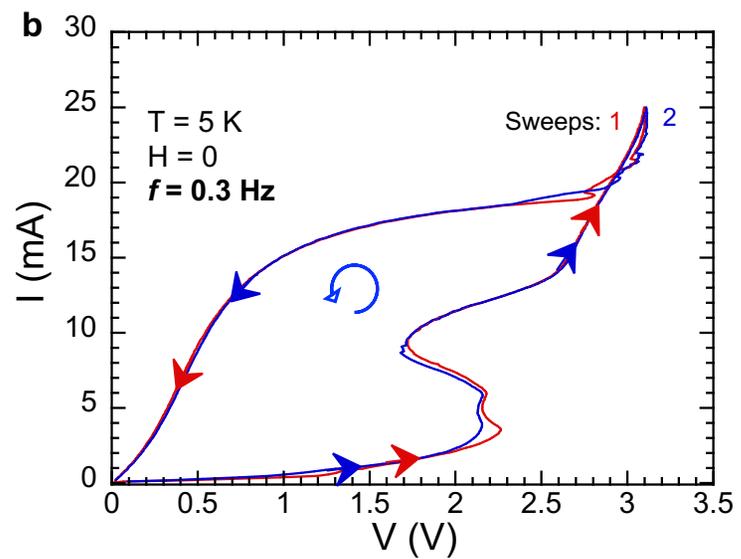

Figure 3

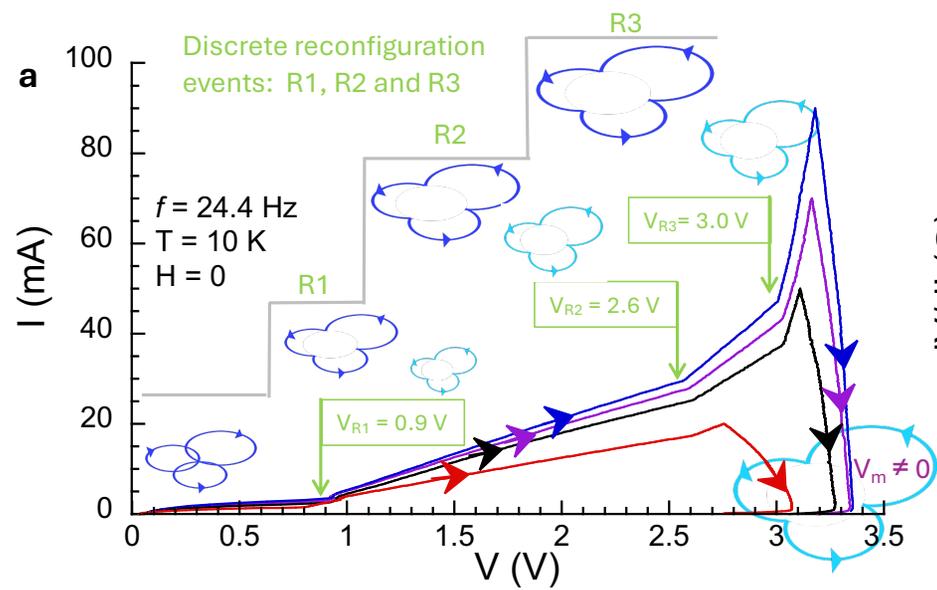
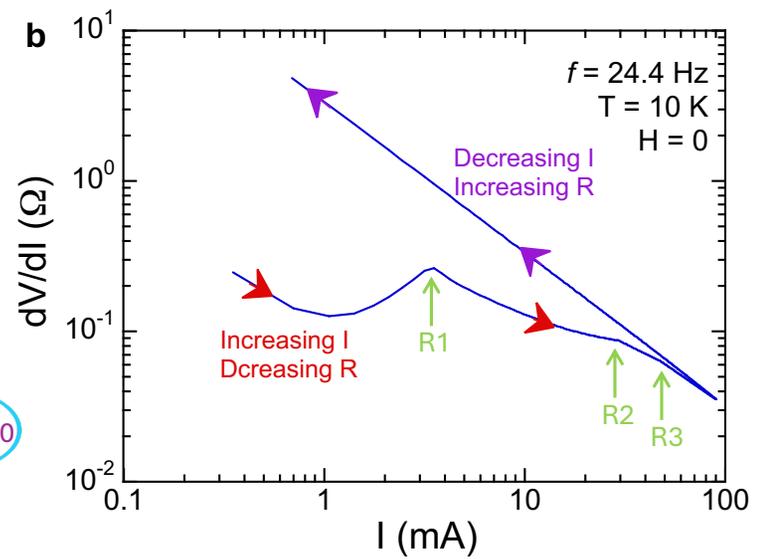

Figure 4

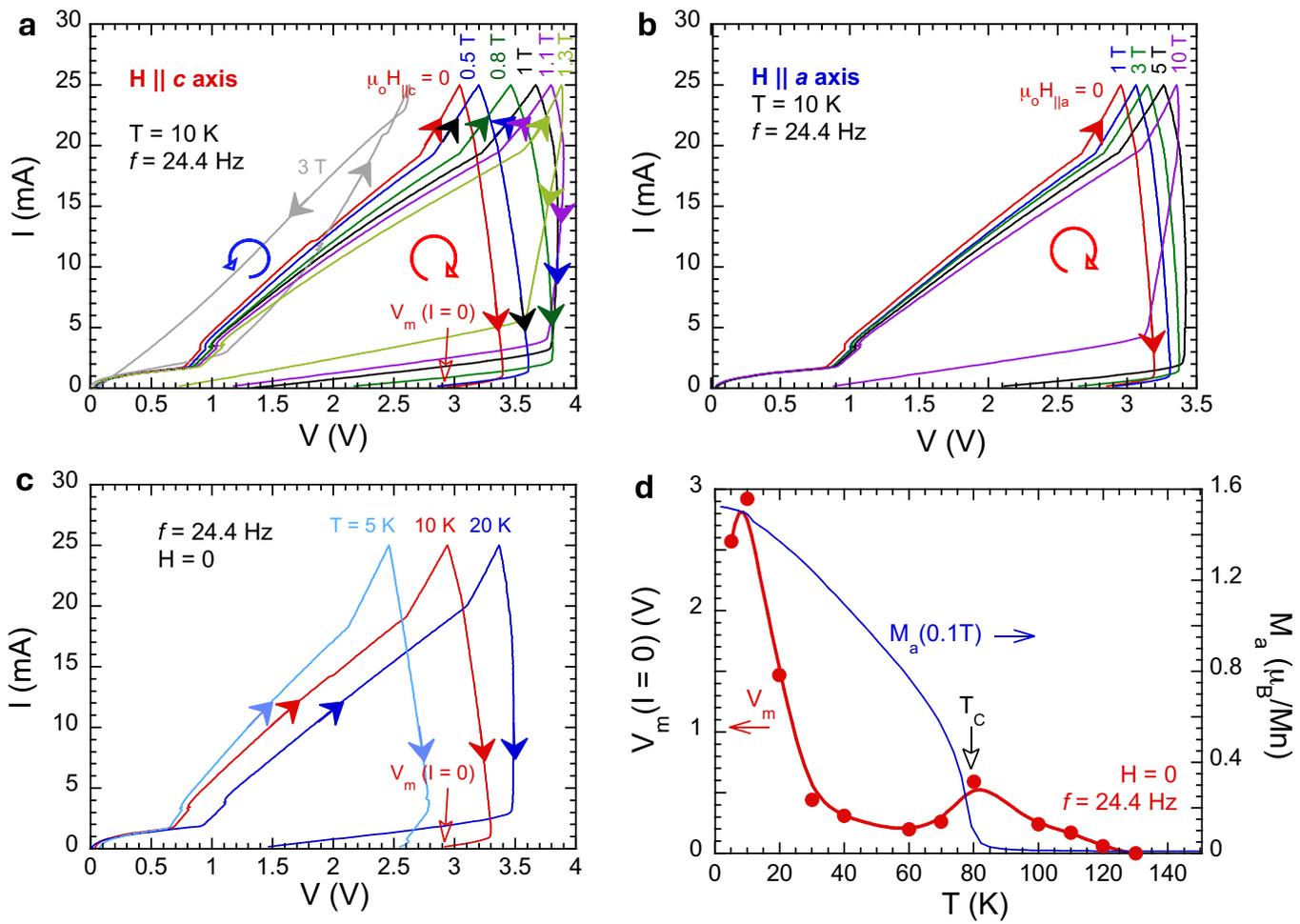

Figure 5

**Inductance Meets Memory in a Quantum Magnet**

Tristan R. Cao[1], Gabriel Schebel[1], Arabella Quane[1], Hengdi Zhao[1], Yu Zhang[1], Feng Ye[2], Longji Cui[3], and Gang Cao[1*]

[1]Department of Physics, University of Colorado at Boulder, Boulder, CO 80309, USA

[2]Neutron Scattering Division, Oak Ridge National Laboratory, Oak Ridge, Tennessee 37831, USA

[3]Department of Mechanical Engineering, University of Colorado at Boulder, Boulder, CO 80309, USA

*Corresponding author: gang.cao@colorado.edu

**Supplementary Note 1**

**Magnetic Structure**

With trigonal symmetry (P-31c), $Mn_3Si_2Te_6$ ferrimagnetically orders at a transition temperature $T_C$ = 78 K with an easy $a$ axis, and a hard $c$ axis. It features a noncollinear magnetic structure with the magnetic space group $C2'/c'$, where the Mn spins lie predominantly within the $ab$ plane with a 10°-tilting toward the $c$ axis in ambient conditions. The noncollinear magnetic structure simultaneously breaks mirror and time reversal symmetries, which is essential for the COC to form below $T_C$. The COC circulate on the edges of $MnTe_6$ octahedra but predominantly within the $ab$ plane and therefore generate the $c$-axis orbital moments.

**Supplementary Note 2**

**No Capacitive Effect**

The capacitance, $C_p$, and the associated dissipation factor, DF, were measured using a QuanTech LCR meter. **SFig. 4a** illustrates $C_p$ and DF as a function of magnetic field H ‖ c axis at T = 3 K and frequency $f$ = 2 MHz. The behavior of DF exhibits significant oscillations with changes in H drastically oscillates with H, which is atypical of a capacitor. DF is defined as DF = R/X where capacitive reactance X = $1/2\pi f C_p$. Typically, DF is expected to increase with $f$. However, in this case, DF actually decreases with increasing $f$, as shown in **SFig. 4b**. It is also noteworthy that DF diverges at low $f$, indicating that electrons are mobile and the charge polarization necessary for capacitor function is not being established. More generally, $Mn_3Si_2Te_6$ becomes highly conductive at H ‖ c axis, which makes any capacitive effects impossible.

**Supplementary Note 3**

**No Joule Heating Induced Artifacts**

A recent PRL *Letter* (https://doi.org/10.1103/ry2d-bgy) attributes the nonlinear transport and bistable switching we reported in $Mn_3Si_2Te_6$ [1] to Joule heating. This conclusion is incompatible with our previously published data and new measurements [1-5].

Most importantly, the *Letter's* measurements were performed at current densities on the order of **$10^3$–$10^4$ A/cm²**, where significant Joule heating is expected. In contrast, our measurements were conducted at **~1 A/cm²** [1-5], several orders of magnitude lower, where direct thermometry confirms negligible temperature rise (< 5 K) [3]. Thermal effects that may occur at extreme current densities cannot be used to interpret our results obtained in this much lower-current regime, underscoring the need for a distinct interpretation of our results.

Joule heating is a gradual, isotropic process, independent of current direction or field orientation. In contrast, our experiments reveal abrupt, first-order transitions that are highly nonlinear, anisotropic, and field- and current-direction dependent – clear signatures that cannot be explained by simple thermal activation.

Because phenomena arising from chiral orbital currents (COC) represent a rapidly developing and impactful research area, it is important to clarify this distinction so that future discussions of current-induced transport in $Mn_3Si_2Te_6$ are guided by a mechanism consistent with the full body of experimental evidence.

In the following, we present multiple, independent lines of evidence, including pulsed measurements, anisotropic field response, power scaling, differential resistance measurements and direct temperature measurements with a sample-mounted Cernox sensor, to demonstrate that the novel phenomena arise from intrinsic COC, conclusively eliminating Joule heating as their origin [1-5].

**Evidence Against Joule Heating**

1. **Bistable Switching and Field Orientation** [1]

    In **SFig. 5**, tiny current steps ($\Delta I \approx 0.005$-$0.01$ mA) trigger discrete voltage jumps ($\Delta V$ up to 0.99 V) that depend strongly on field orientation: Robust for H || $c$ (**SFig. 5b**) but absent for H || $a$, where Ohmic behavior returns (**SFig.5c**). ***If this switching were due to Joule heating, V or R would decrease rather than increase because R at H = 0 follows an <u>insulating behavior</u>*** (***i.e., the higher temperature corresponds to lower resistance***). Moreover, the switching is between two values of V, thus the bistable switching essentially independent of applied currents [1]. Joule heating, being isotropic and continuous, cannot produce such switching.

2. **Nonlinearity in Differential Resistance**

In **SFig. 6**, differential resistance dV/dI vs DC current up to 30 mA at $f$ = 1 Hz with a 15 mA AC modulation shows pronounced, symmetric nonlinearity about I = 0 and sharp features near well-defined threshold currents, which are clear signatures of intrinsic, current-driven reconfiguration of orbital textures. Importantly, no monotonic resistance drop or runaway behavior is observed at high bias, ruling out Joule-heating artifacts. The response is fully reversible and frequency dependent. In particular, the strongly nonlinear yet symmetric dV/dI curves are incompatible with heating, which would produce a smooth, monotonic resistance decrease with increasing current.

3. **Power Inversely Proportional to Current – Current-Direction-Dependent Resistance** [5]

   In **SFig. 7**, the resistance R remains essentially unchanged and small (~ 0.001 Ohm) with increasing current I but rises rapidly by nearly three-orders of magnitude ***only when current is reduced***, and power dissipation increases as current decreases: e.g., 2.0 mW at I = 20 mA vs. 4.3 mW at I = 4 mA. This is clearly inconsistent with Joule heating: Joule heating, which scales with $I^2R$, would rise when the current I increases, and fall when the current I decreases.

4. **Extreme Field Anisotropy** [1,5]

   In **SFig. 8** I-V curves at H || $a$ and H || $c$ should behave similarly, independent of the field orientation if Joule heating were a driving force. But the I-V curves in **SFig. 8** respond vastly differently to different field orientations (see red curve for H || $c$ and blue curve for H || $a$). At I = 2 mA, power differs by three orders of magnitude between H || $c$ and H || $a$ (0.008 mW vs. 2.6 mW), which cannot arise from an isotropic heating mechanism.

5. **Hysteresis Loop Reversal of I-V Curves** [5]

   Similarly, in **SFig. 9**, the I-V curves at H || $a$ and H || $c$ respond entirely differently to different field orientations: H || $c$ causes a ***clockwise loop*** indicating the emergent inductive behavior

[5], i.e., increasing current I only slight changes voltage V, whereas decreasing current I induces an increase in V. In contrast, H∥ *a* results in a ***counterclockwise loop***: increasing I causes two negative differential resistance transitions while decreasing I leads to a decrease in V (see red curve for H ∥ *c* and blue curve for H ∥ *a*). The contrasting behaviors for H ∥ *a* and H ∥ *c* are once again inconsistent with Joule heating.

6. **Pulsed vs. Continuous Current Measurements** [5]

   In **SFig. 10**, I-V curves under pulsed (0.1 s on/off) and DC current are nearly identical up to 12 mA, confirming that thermal accumulation plays no role.

7. **Direct Thermometry** [3]

Cernox-based measurements with a **Cernox thermometer attached to a single-crystal sample (SFig.11)** show the increase of temperature ΔT < 5 K even at the highest currents used and decreasing ΔT with increasing field, incompatible with thermal runaway (**SFigs. 12-13**). Note that ***the current density is on the order of 1 A/cm$^2$*** [3].

**Conclusion**

Our comprehensive data, spanning anisotropic field dependence, first-order switching, inverse power scaling, hysteretic I-V loops, pulsed/DC equivalence, and direct thermometry, conclusively rule out Joule heating as the mechanism for the nonlinear transport in Mn$_3$Si$_2$Te$_6$ [1-5]. These effects originate from reconfigurable COC domains, representing a new type of collective electronic state [5].

Because COC phenomena are a rapidly developing frontier, establishing the correct microscopic mechanism is essential for informing future experimental and theoretical efforts. We offer this clarification with the aim of fostering clearer distinctions between intrinsic chiral-orbital-current phenomena and thermal artifacts.

**Supplementary References**

**Supplementary Figure Captions**

**SFig.1. Multi-sweep I-V curves revealing the system retaining memory of the prior cycle. a, b,** Multi-sweep I-V curves at 24 Hz at T = 10 and 80 K showing state carry-over. Note the effect gets saturated when sweeps are greater than 5 times.

**SFig.2. Magnetic-field dependence and orientation selectivity.** Comparison of $V_m$ vs H for both orientations with $a$-axis magnetoresistance ratio $[\rho_a(H)-\rho_a(0)]/\rho_a(0)$, highlighting the competition between coherent and disordered COC states.

**SFig.3. Temperature dependence and persistence of memory above $T_C$. a, b**, I-V loops at representative temperatures from 40 K to 130 K, H = 0, and $f$ = 24.4 Hz, showing robust clockwise inductive loops and finite $V_m$ well into the paramagnetic state.

**SFig. 4. Capacitance $C_p$ and dissipation factor DF for x = 0. a,** $C_p$ and DF at T = 3 K and $f$ = 2 MHz as a function of magnetic field H aligned along the c axis. b, DF at T = 10 K and H = 0 as a function of $f$. Note that DF diverges as $f$ approaches zero, suggesting the absence of charge polarization needed for a capacitor.

**SFig. 5. Time-dependent bistable switching [1]:** The $a$-axis voltage $V_a$ as a function of time t at T = 10 K for **(a)** H = 0, **(b)** $\mu_oH_{\|c}$ = 7 T and **(c)** $\mu_oH_{\|a}$ = 7 T [1]. Note that there are only two values of V independent of applied currents I.

**SFig.6. Nonlinearity in Differential Resistance:** dV/dI vs DC current up to 30 mA at $f$ = 1 Hz with 15mA AC modulation shows pronounced, symmetric nonlinearity about I=0I=0 with sharp features near threshold currents, signatures of intrinsic current-driven reconfiguration of orbital textures. Crucially, no monotonic resistance decrease or runaway behavior is observed at high bias, ruling out Joule-heating artifacts. The nonlinearity is reversible and frequency dependent, consistent with electronic origin rather than thermal effects.

**SFig. 7. Resistance R rises with reducing current I [5]:** R remains essentially unchanged with increasing current but rises rapidly by nearly three-orders of magnitude only when current is reduced at $\mu_oH_{\|c}$ = 14 T and 5 K. Note that resistance R depends on current direction.

**SFig. 8. Highly anisotropic I-V curves [1,5]:** The I-V curves driven by DC at H = 0 T (black), $\mu_oH_{\|c}$ = 14 T(red) and $\mu_oH_{\|a}$ = 14 T (blue). Note the high sensitivity of I-V curves to field orientation [1,5].

**SFig. 9. Highly anisotropic I-V curves [5]:** The I-V curves at $f$ = 1.0 Hz and 20 K for $\mu_oH_{\|c}$ = 5 T (red) and $\mu_oH_{\|a}$ = 5 T (blue). Note that the change of the I-V loop orientation due to the change of field orientation [5].

**SFig. 10. Comparison of I-V curves measured under continuous DC (black) and pulsed (red) current excitation at 0 T [5].** Pulsed data were collected using a 0.1 s on / 0.1 s off duty cycle. The nearly overlap of the curves up to 12 mA demonstrates that Joule heating is negligible under the measurement conditions.

**SFig. 11. Direct Temperature Measurements with a Sample-Mounted Cernox Sensor [3]:** A *Cernox thermometer* (gold, front) is ***thermally contacted with a single-crystal sample*** of $Mn_3Si_2Te_6$ (black, behind the Cernox). The thin gold wires are electrical leads electrically attached to the sample and Cernox with an EPO-TEK H20E epoxy (silver). The light brown background is GE varnish used to thermally anchor the Cernox and the sample. Note that H||c-axis and I || a-axis.

**SFig. 12. Temperature change ΔT (K) as functions of applied current I (mA) and current density J (A/cm²)** at T = 30 K for representative magnetic fields $\mu_oH \| c$ axis [3].

**SFig. 13. Temperature change ΔT (K) as functions of applied current I (mA) and current density J (A/cm²)** at $\mu_oH_{\|c}$ = 3 T for representative temperatures T [3].

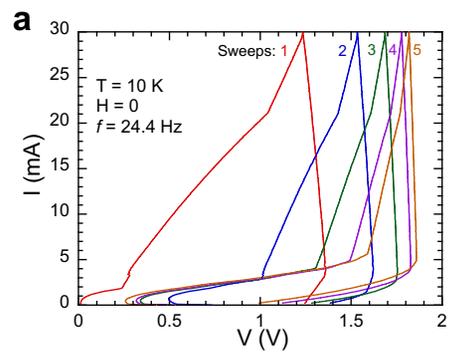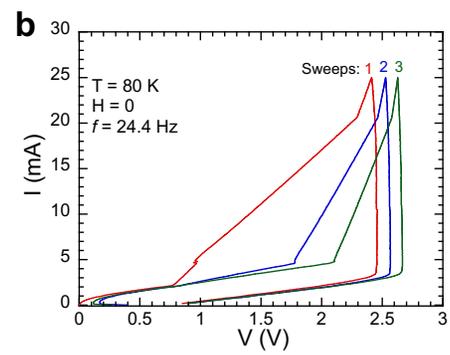

SFigure 1

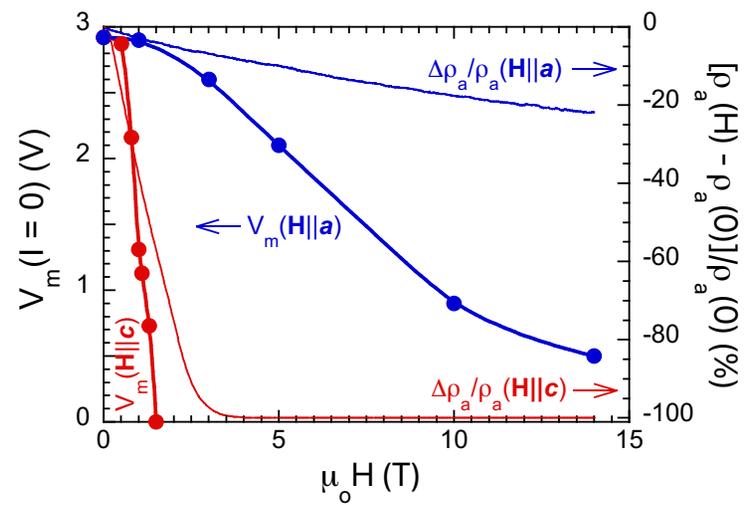

SFigure 2

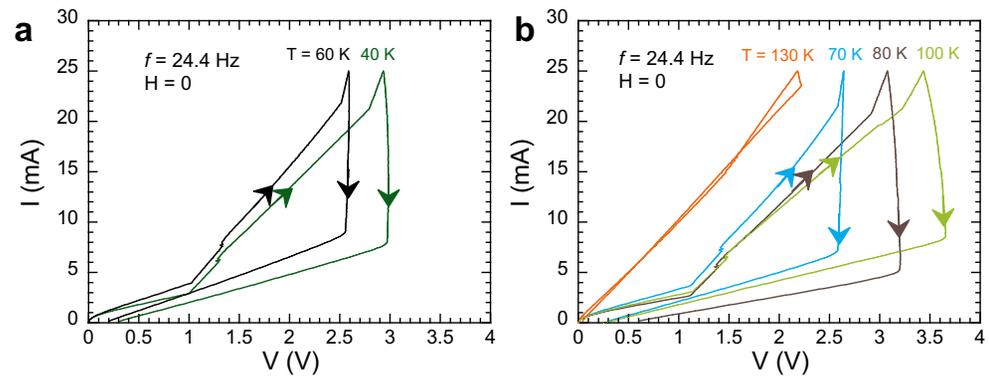

SFigure 3

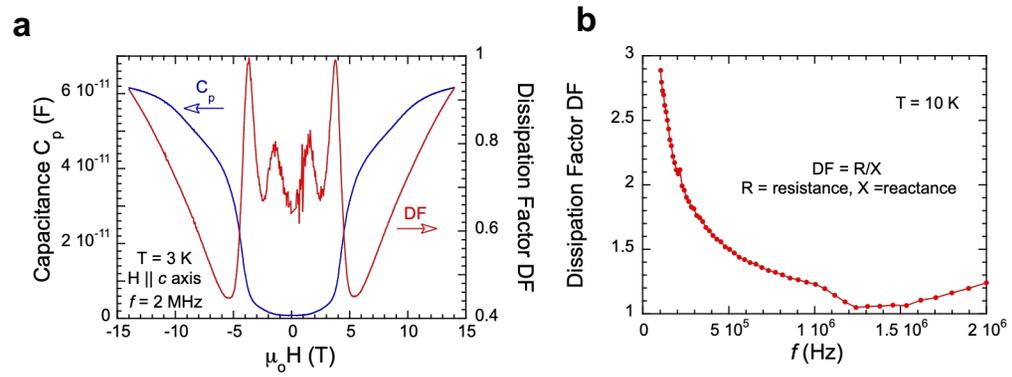

SFigure 4

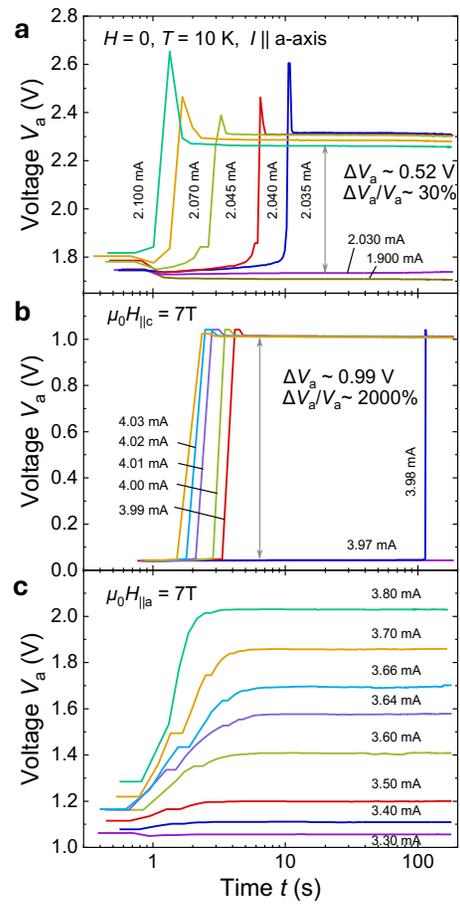

SFigure 5

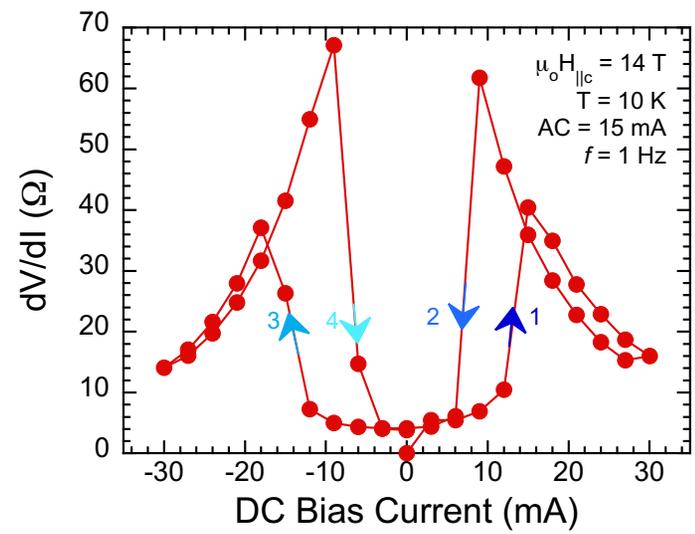

SFigure 6

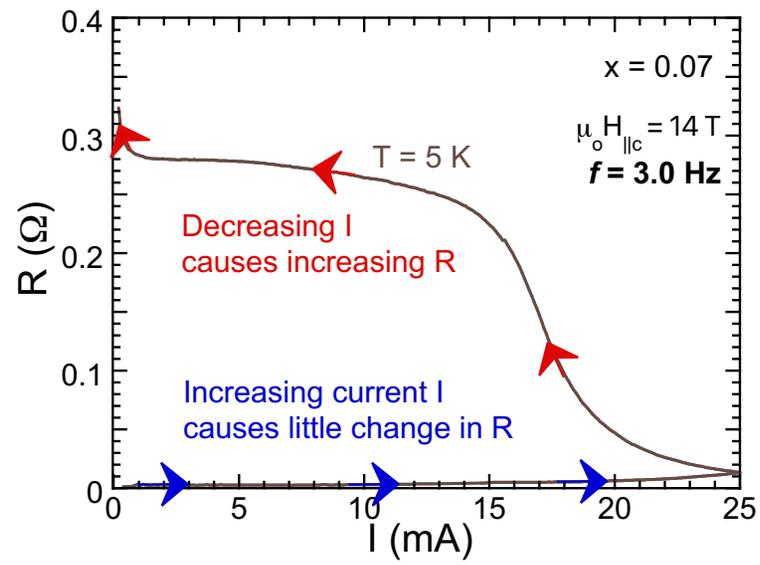

SFigure 7

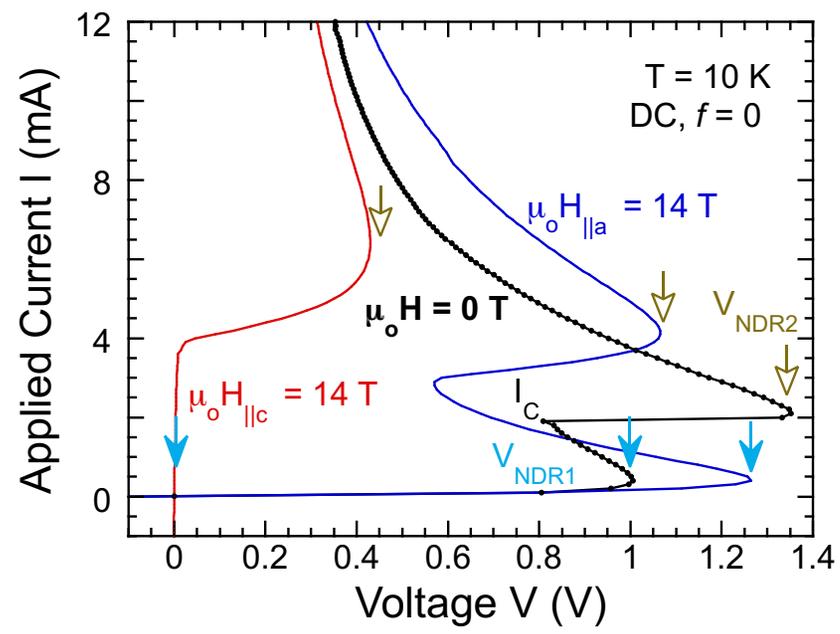

SFigure 8

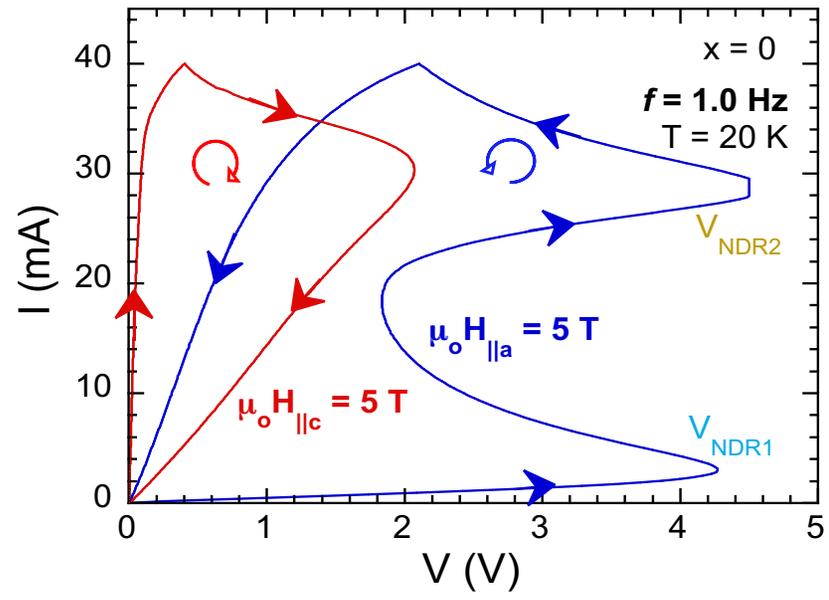

SFigure 9

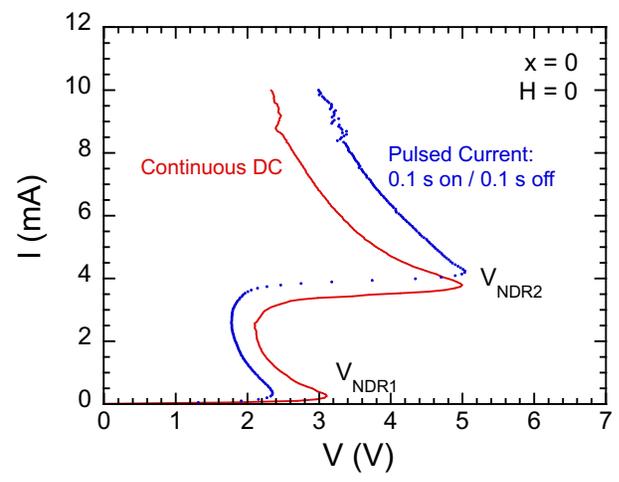

SFigure 10

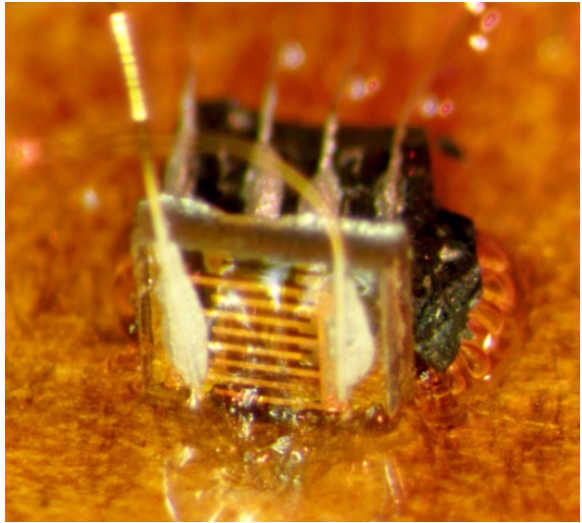

SFigure 11

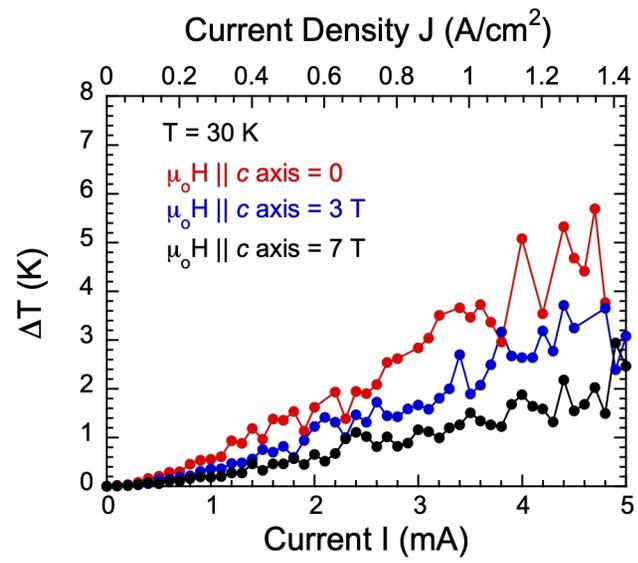

SFigure 12

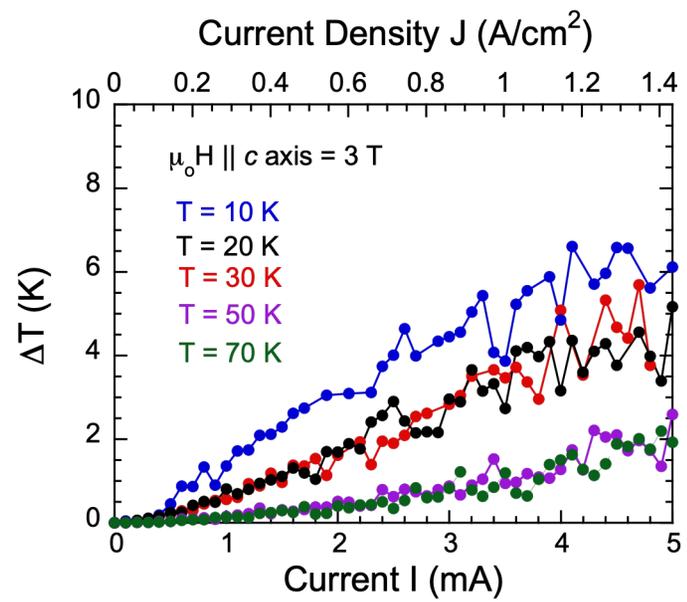

SFigure 13